\documentclass[aps,pra,twocolumn]{revtex4-1}
\usepackage[utf8]{inputenc}
\usepackage{graphicx}
\usepackage{amsmath}\usepackage{amssymb}\usepackage{epsfig}

\begin{document}
\title{Spontaneous formation of Kagomé lattice in two-dimensional Rydberg atoms}
\author{Yu. E. Lozovik$^{1,3}$, I. L. Kurbakov$^1$, and G. E. Astrakharchik$^2$}
\affiliation{$^1$Institute for Spectroscopy RAS, 142190 Troitsk, Moscow, Russia}
\affiliation{$^2$Departament de F\'{\i}sica, Campus Nord
B4-B5, Universitat Polit\`ecnica de Catalunya, E-08034 Barcelona, Spain}
\affiliation{$^3$MIEM, National Research University Higher School of Economics, 101000 Moscow, Russia}
\date{December 23, 2019}
\begin{abstract}
Two-dimensional Rydberg atoms are modeled at low temperatures by means of the classical Monte Carlo method. The Coulomb repulsion of charged ions competing with the repulsive van der Waals long-range tail is modeled by a number of interaction potentials. We find that under specific conditions the usual triangular crystal becomes unstable with respect to more exotic lattices with Kagomé, flower, molecular and rectangular-chain packing. Ground-state configurations are obtained by means of the annealing procedure and their stability is additionally studied by the normal modes analysis. While commonly the square lattice is mechanically unstable due to softening of the shear modulus, we were able to find a specific set of parameters for which the square lattice can be made stable.
\end{abstract}
\maketitle

\section{Introduction}

The question of the optimal packing is a long-standing problem which often cannot be easily solved. Historically, the greatest minds, including Gauss and Kepler have been working on what is the maximal density packing of canon balls. The Kepler's conjecture\cite{Kepler1611,Kepler1966english} dating back to 1611 states that the face-centred cubic (fcc) and hexagonal-close packed (hcp) crystal packing have the largest average density in three dimensions. While Gauss proved the conjecture for the case of regular lattice arrangement already in year 1831, only very recently the general proof involving also non-regular arrangements became available\cite{arxiv1501.02155,KeplerConjecture2017}. A related question arises in finding the phase diagram of solids\cite{KittelBook} where different packings are possible, including fcc, hcp, and body-centered cubic (bcc) structures. Although being conceptually simple, the question of finding the configuration with the lowest energy turns out to be numerically quite complicated as the differences in the energy between possible packing symmetries can be extremely small.

On the other hand, in two dimensions (2D) the situation is much simpler as in this case commonly only one crystal packing is possible. The hexagonal (triangular) lattice (Abrikosov lattice\cite{Abrikosov1957} in the field of superconductors) typically has the smallest energy while the square lattice typically suffers a mechanical instability due to softening of the shear modulus. Hence, it is highly unusual to find a purely repulsive two-dimensional single-component system in which a packing different from triangular one is realized in the ground state.

The interaction potential $U(r)$ between Rydberg atoms contains a repulsive van-der Waals $1/r^6$ long-range tail while the screening of excitations at short distances is usually modeled by a plateau. Physically, the shape of the interaction potential at intermediate distances is determined by the properties of the wave function of the highly excited electron in a Rydberg atom. A common phenomenological model\cite{PhysRevLett.104.195302,PhysRevLett.108.265301,arxiv1606.04267} approximates the interaction potential by $U(r) = C_6 /(r^6 + a^6)$, for which atomic clustering\cite{pre092052307} was predicted in a classical system. This particular choice of the shape of the interaction potential does not take into account the Coulomb repulsion of the charged ions inside the atomic core: $U(r) \propto 1/r$ for $r\ll a$. Purely algebraic $1/r^6$ interaction potential results in a ``standard'' phase diagram with transitions between gas (superfluid and normal) and solid phases\cite{Osychenko2011}. Instead, inclusion of the short-range shoulder might lead to highly unusual properties. The dipole-blockaded ($1/r^3$ with short range plateau) interactions were shown to result in formation of a crystal of mesoscopic superfluid droplets\cite{PhysRevLett.105.135301}. Soft shoulder potentials might form glassy phases with a simultaneous finite superfluid fraction\cite{PhysRevLett.116.135303}.

In this paper, we study a system of Rydberg atoms interacting via repulsive van-der Waals $1/r^6$ potential taking into account the Coulomb repulsion and show its decisive role in changing entirely the structure of the ground state of the 2D classical system. We find that under certain specific conditions, the usual triangular lattice is no longer realized in the ground state of the system. Instead, we find that atoms spontaneously form Kagomé and a number of other exotic lattices.

\section{Model and methods}

We perform a classical Monte Carlo simulation of Rydberg atoms in a two-dimensional geometry. The following two models for the pair interaction potential $U(r)$ are used,
\begin{equation}\label{U1}
U_1(r)=\frac{C_6}{r/a+(r/a)^6},
\end{equation}
\begin{equation}\label{U2}
U_2(r)=C_6\frac{1+a/r}{1+(r/a)^6},
\end{equation}
where the coefficient $C_6>0$ is positive. In both models the Coulomb $1/r$ short-range repulsion is smoothly matched with van der Waals $1/r^6$ long-range asymptotic, with the crossover occurring at distances of the order of $r\approx a$.

The guiding parameters of the problem are (i) the dimensionless density $na^2$ and (ii) the dimensionless temperature $T/C_6$. In the following we use notation in which the Boltzmann constant is $k_B=1$. We are interested in the ground state properties corresponding to $T=+0$.

The simulations are performed in a rectangular box of size $L_x\times L_y$ with periodic boundary conditions. The aspect ratio $L_x/L_y$ is commensurate with the elementary cell of the crystal. The search for the ground-state configuration is made for $N\sim 1000$ particles which in 2D is a sufficiently large number to define correctly the ground-state configuration. An annealing method is used in which the temperature is lowered from an initial high temperature ($T/C_6 = 1$) to a small one ($T/C_6\sim10^{-6}$) at a very slow exponential rate. The actual value of the box aspect ratio, $L_x/L_y$, is selected from the minimum energy condition. Instability of a certain type of lattice is determined by its collapse in the annealing procedure and is additionally confirmed by the normal modes analysis. In the cases, where different metastable packings are possible, we compare the energies of the different geometries.

\section{Results}

Our main result consists in observation of spontaneous formation of the Kagomé and a number of other lattices in the ground state in a wide range of densities, instead of the usual triangular lattice.

As the temperature is decreased, the potential energy starts to dominate over the kinetic (thermal) energy. In a classical system this leads to a phase transition between a state with a translational symmetry (a gas or a liquid) to a state with broken translational symmetry (crystal).

\begin{figure}
\centering
\includegraphics[width=\columnwidth]{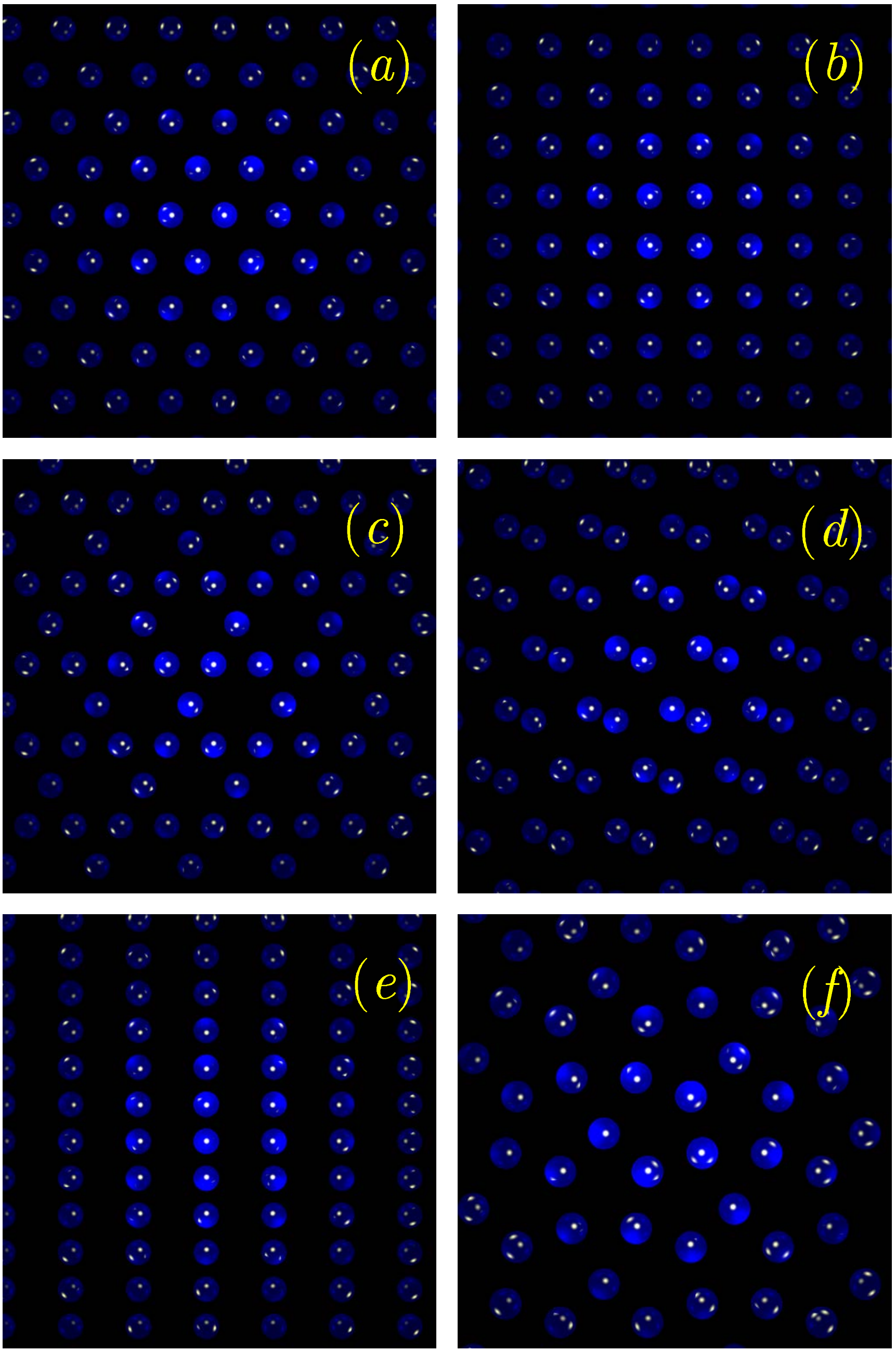}
\caption{Characteristic examples of packing structures of the ground-state crystal obtained for potentials~(\ref{U1})-(\ref{U2}):
(a) usual triangular lattice, (b) square lattice, (c) Kagomé lattice, (d) molecular crystal, (e) chain crystal, (f) flower lattice.}
\label{Fig1}
\end{figure}

Figure~\ref{Fig1} provides characteristic examples of the ground-state packing. The usual triangular lattice, shown in Fig.~\ref{Fig1}a at $na^2=4$ for potential~(\ref{U2}), is the most common structure. To our best knowledge this is the only ground-state packing known in single-component systems with monotonously decaying two-body potential.

Figure~\ref{Fig1}c shows Kagomé packing which we find for interaction potential~(\ref{U2}) at density $na^2=2.56$. The lattice consists of vertices and edges of the trihexagonal tiling \cite{Mekata2003}. For the considered parameters, the energy of Kagomé lattice is lower than the energy of the triangular lattice by $0.9\%$. It should be noted that it is rather common\cite{RevModPhys.89.035003} that the energy difference between different packing configurations is extremely small, nevertheless the system naturally chooses the state with the minimal energy. For exactly the same choice parameters, it turns out that the triangular lattice is completely unstable while the Kagomé lattice is stable up to the critical temperature $T_c/C_6=0.038(2)$ and melts for higher temperatures. The modified Lindemann ratio\cite{pla109000289} at transition point equals to $\gamma_L=0.173(6)$ which is consistent with the value found in other classical 2D crystals.

Figure~\ref{Fig1}f shows a more intricate lattice which we obtain for potential~(\ref{U1}) at density $na^2=1.96$. We coin it a ``flower'' packing for the reasons which will become evident later. This packing has energy lower by 0.32\% than that in the triangular lattice. We estimate the melting temperature as low as $T_c/C_6=0.0041(2)$ while the triangular lattice is completely unstable and melts at any temperature.

In addition, other non-trivial orderings are possible. For potential~(\ref{U2}) at density $na^2=1.21$ we find an isosceles triangular lattice of atom pairs which is shown in Fig.~\ref{Fig1}d. It can be interpreted as crystal composed of molecules. As well we succeeded in localizing a rectangular lattice shown in Fig.~\ref{Fig1}e which can be interpreted as two-dimensional system of chains. This packing is obtained for potential~(\ref{U2}) at $na^2=1.44$. The ground-state energies of the molecular crystal and system of chains lower than the energy of the triangular lattice by $4.3\%$ and $6.4\%$, respectively.

Probably, the most striking ground-state configuration is a {\it stable} square lattice shown in Fig.~\ref{Fig1}b. It is observed for potential~(\ref{U1}) at density $na^2=1.44$. Its energy is lower by $0.3\%$ as compared to that of a triangular lattice. The melting temperature of the square lattice is found to be as low as $T_c/C_6=0.0067(1)$. The corresponding modified Lindemann ratio is equal to $\gamma_L=0.152(2)$. Whereas, for the same choice of parameters triangular lattice is unstable. The fact of stability of the square lattice is a feature of the transient regime between the short-range $1/r$ and long-range $1/r^6$ asymptotic in the interaction potential~(\ref{U1}). Indeed, for both Coulomb $1/r$\cite{prb015001959} and pure $1/r^6$\cite{Osychenko2011} potentials a triangular lattice is a stable ground-state configuration whereas a square lattice is unstable.

An additional information of the structure of an elementary cell can be obtained from the static structure factor which quantifies correlations in momentum space,
\begin{equation}
S({\bf k}_2 - {\bf k}_1) = (1/N)\langle \rho_{{\bf k}_1}\rho_{{\bf k}_2}\rangle,
\label{Sk}
\end{equation}
where $\rho_{\bf k} = \sum_{j=1}^N e^{i{\bf kr}_j}$ is the Fourier transform of the density operator.

\begin{figure}
\centering
\includegraphics[width=\columnwidth]{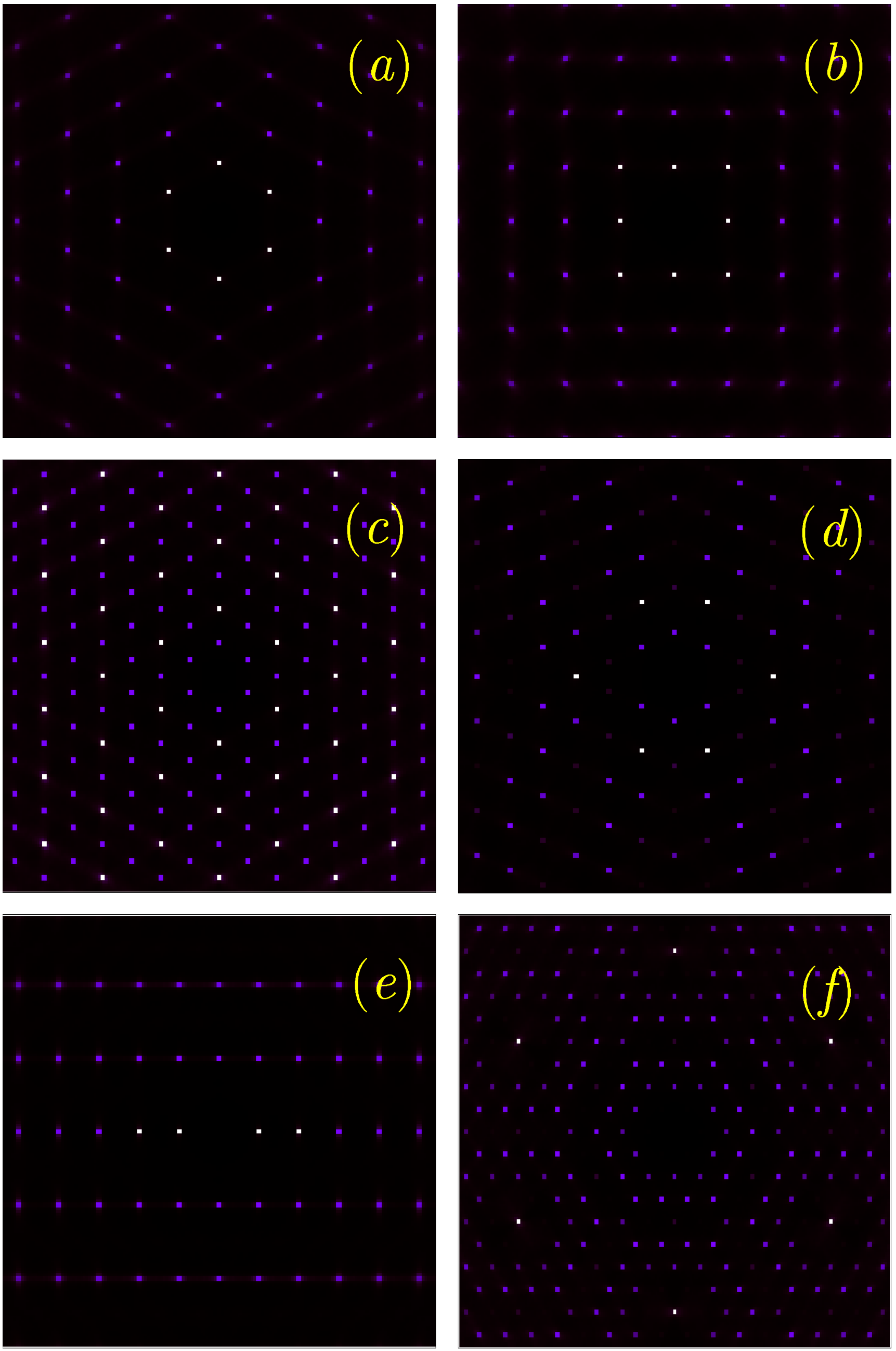}
\caption{Static structure factor~(\ref{Sk}) which quantifies the correlations in momentum space for the lattices shown in Fig.~\ref{Fig1}.}
\label{Fig2}
\end{figure}

Figure~\ref{Fig2} reports the static structure factor for different packings. The height and the width of the peaks depend on the exact temperature and density, while the general structure remains the same for a given type of packing. It can be seen that the triangular lattice (Fig.~\ref{Fig2}a) has $\pi/3$ symmetry while the square lattice is invariant under $\pi/2$ rotation. Kagomé lattice (Fig.~\ref{Fig1}c) clearly shows the period doubling in the reciprocal space as compared to the standard triangular lattice (Fig.~\ref{Fig1}a). Of particular appeal is the flower lattice. In coordinate space (see Fig.~\ref{Fig1}f), it might seem that it should possess a pentagonal ordering with no diagonal long-range order. But this is not so, as a set of sharp peaks are visible in the reciprocal space (Fig.~\ref{Fig1}f) which means that a diagonal long-range order is present. The positions of the peaks form a centered-hexagonal periodic structure which might be found aesthetically beautiful due to its resemblance with a flower (hence the name we give to this lattice).

\section{Discussion}

All observed exotic lattices appear for the density parameter of the order of unity, $na^2\sim 1$. This means that at this conditions neither the long-wavelength $1/r^6$ nor the short-wavelength $1/r$ asymptotics are dominant. Therefore, the properties of the ground state depend on the mesoscopic part of the potential.

The specific mesoscopic behavior of the interaction potential depends on the wave functions of the electron on the outer shell. Its precise calculation is rather complicated and can be done, for example, within the framework of density functional theory (DFT) \cite{DFT}. In order to avoid the difficult task of calculating the specific behavior at $r\sim a$ as well as to make our results more general we consider several model potentials~(\ref{U1}-\ref{U2}), all of them possessing the correct short- and long-range asymptotics, and a general class of potentials described by the following model,
\begin{equation}
U_{\rm gen}(r)=C_6\frac{A+a/r}{1+(r/a)^l},
\label{Ugen}
\end{equation}
where $A>0$ is a positive offset and $l\gg 1$ is the exponent of the long-range power law decay. For a general class of potentials described by~(\ref{Ugen}), we found the same lattice types as that we had obtained for potentials~(\ref{U1}) and~(\ref {U2}), which we report in Figs~\ref{Fig1}-\ref{Fig2}.

\begin{figure}
\centering
\includegraphics[width=0.4\textwidth]{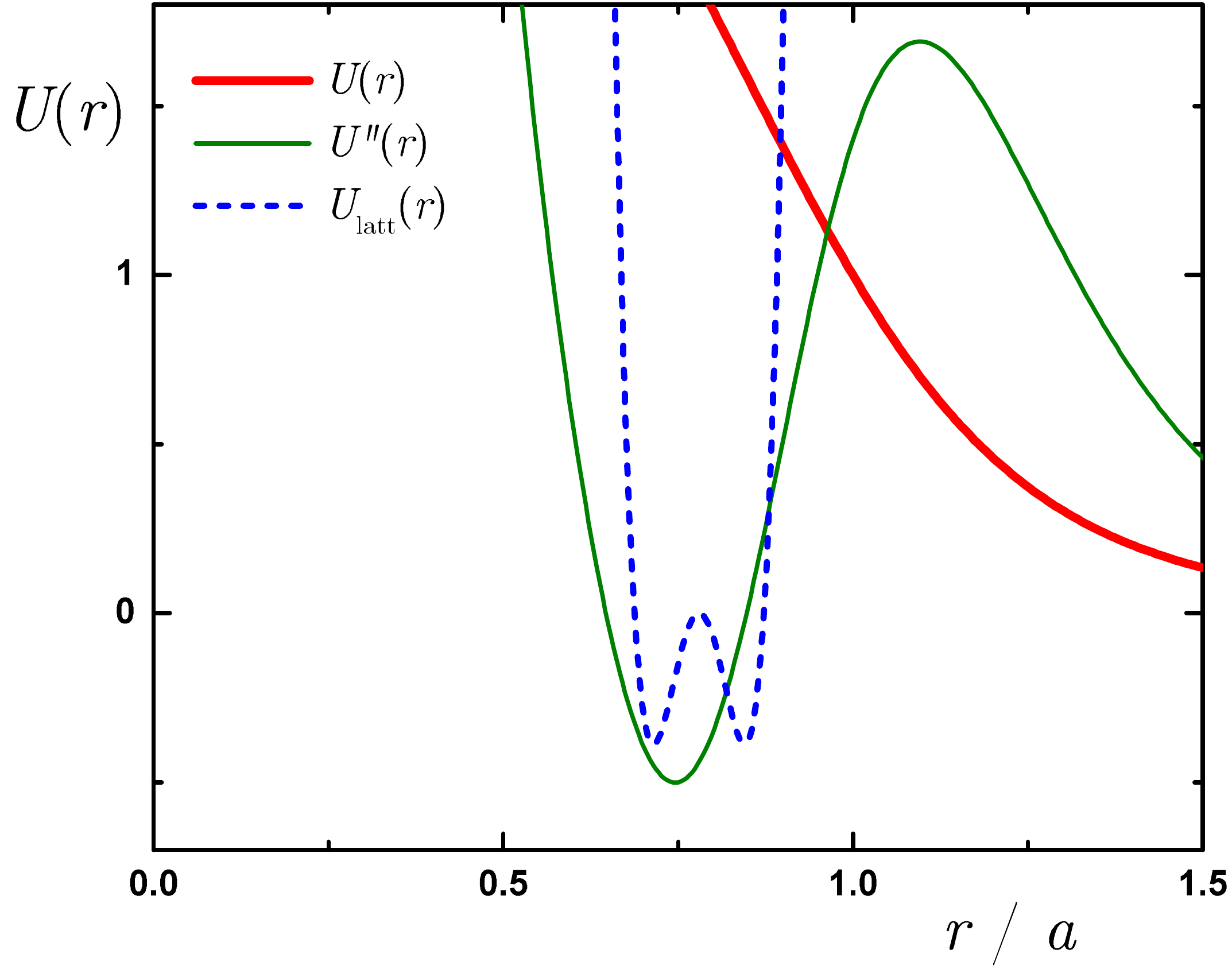}
\caption{
The thick red line shows the potential $U_2(r)$, the thin green line is its second derivative $U_2''(r)$ ($\times 0.2$), the dashed blue line is the potential energy $U_{\rm latt}(r)$ ($\times 500$) of a single atom in a perfect one-dimensional lattice, Eq.~(\ref{UU}) at $R_{\rm nn} = 0.78a$.
The instability is clearly visible: at the interparticle distance $r = R_{\rm nn}$, the potential~(\ref{UU}) has a maximum instead of a minimum (the derivative $U_2''(0.78a)$ is negative). The presence of two potential minima~(\ref{UU}) at $r = 0.71a$ and $r = 0.85a$ leads to a non-equidistant arrangement.}
\label{Fig3}
\end{figure}

It is important to understand how it could have happened that for a centrally symmetric repulsive monotonically decreasing interaction potential, the triangular lattice suddenly turned out to be unstable.
`
For the sake of simplicity, we first consider the one-dimensional case, using the potential~(\ref{U2}) for being specific. Figure~\ref{Fig3} shows the spatial dependence of that potential together with its second derivative. It might happen that for certain densities the second derivative of the $U_2''(r)$ pair potential is negative at nearest-neighbor distance $r=R_{\rm nn}$. It is instructive to calculate the potential energy of interaction between a single atom with the rest of the particles forming a 1D perfect crystal. It is given by
\begin{equation}
U_{\rm latt}(r)=\sum_{j\ne0}[(U_2(|r-(j+1)R_{\rm nn}|)-U_2(|jR_{\rm nn})|)],
\label{UU}
\end{equation}
where for convenience we chose an arbitrary additive constant in such a way that the potential energy vanishes for nearest neighbors, $U_{\rm latt}(R_{\rm nn})=0$. Figure~\ref{Fig3} shows an example in which the potential energy~(\ref{UU}) does not have a minimum at $r=R_{\rm nn}$ but rather a maximum. In other words, for an equidistant chain, the harmonic oscillator becomes {\it absolutely unstable} due the second derivative being negative at $r=R_{\rm nn}$. A stable configuration corresponds than to {\it two} interparticle distances: one on the left and the other on the right. As a result, the equidistant arrangement of atoms in a perfect chain becomes unstable with respect to some nonequidistant configuration even at $T=+0$.

\begin{figure}
\centering
\includegraphics[width=\columnwidth]{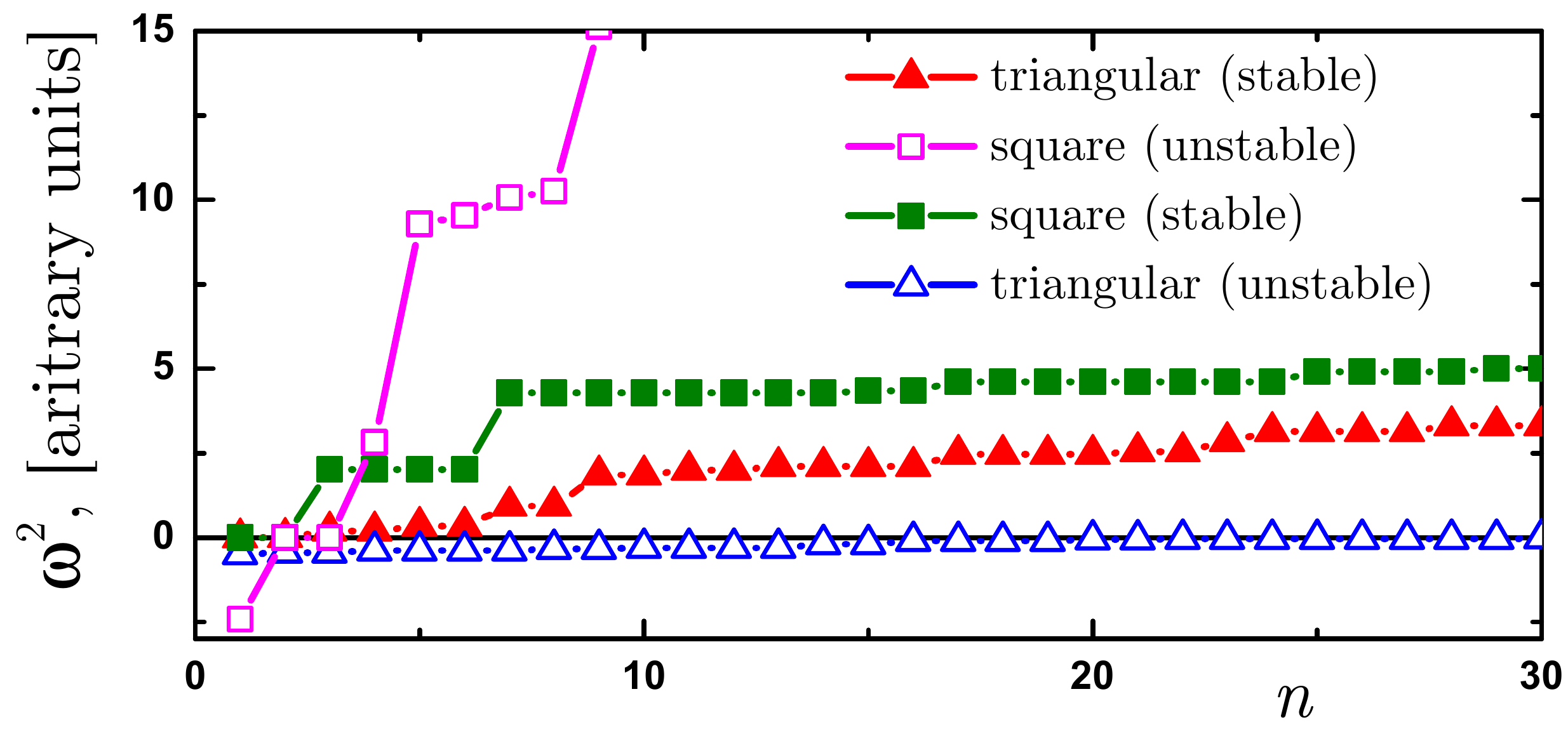}
\caption{Square of the eigenfrequency of the normal modes for triangular ($\triangle$) and square ($\square$) lattices as a function of the eigenmode number. Solid (open) symbols denote stable (unstable) configurations. Stable $\blacktriangle$ and unstable $\square$ lattices are obtained with potential~(\ref{U2}) at $na^2=4$. Stable $\blacksquare$ and unstable $\triangle$ lattices are obtained with potential~(\ref{U1}) at $na^2=1.44$.}
\label{Fig4}
\end{figure}

A similar reasoning can be applied to lattices in the 2D case. If the potential energy of an atom in the field of the remaining particles in a perfect lattice has a maximum instead of a minimum, this lattice configuration is unstable. That is what happens for the triangular lattice for parameters reported in Fig.~\ref{Fig4}.

Finally, we study the mechanical stability of the system by performing the normal modes analyses\cite{Ashcroft76}. As usual, the potential energy of the crystal is expanded in the vicinity of the minimum as a quadratic form $U({\bf r}_1,...{\bf r}_N)=U({\bf r}_1^0,...{\bf r}_N^0)+(1/2)\sum_{jk\alpha\beta}(r_{j\alpha}-r_{j\alpha}^0)H_{j\alpha}^{k\beta}(r_{k\beta}-r_{k\beta}^0)+...$, where indices $j,k=1,2,...N$ perform the summation over particles and $\alpha,\beta=x,y$ over dimensions, and
\begin{equation}
H_{j\alpha}^{k\beta}=\frac{\partial}{\partial r_{j\alpha}^0}\frac{\partial}{\partial r_{k\beta}^0}
U(r_{1x}^0,...r_{Ny}^0),
\label{Hessian}
\end{equation}
is the second-derivative (Hessian) matrix. Frequencies $\omega$ of small-amplitude oscillations in the vicinity of the minimum are obtained from the Newton equations of motion according to the eigenvalue problem
\begin{equation}
{\rm det}\left|\left|H-m\omega^2I\right|\right|=0
\label{omega2}
\end{equation}
for the Hessian~(\ref{Hessian}) with $I_{j\alpha}^{k\beta}=\delta_{jk}\delta_{\alpha\beta}$ being is the unit matrix.

Figure~\ref{Fig4} reports square of the frequency $\omega^2$ for the lowest excitation modes as obtained from the eigenvalue problem~(\ref{omega2}) for the Hessian matrix. In all cases, there are two trivial soft modes, $\omega^2=0$, corresponding to translation of the center of mass along $x$ and $y$ directions. Those modes do not affect the stability of the system. First, it is instructive to consider the typical dependence for the stable triangular lattice. Such a dependence is shown in Fig.~\ref{Fig4} by solid triangles. Apart from the two trivial soft modes all mode are real and the lowest modes correspond to gapless excitations (phonons). Instead, for a usual square lattice shown by open squares in Fig.~\ref{Fig4} (potential~(\ref{U2}) at $na^2=4$), the lowest mode has an imaginary frequency, $\omega^2<0$, due to the system instability to the shear wave. The first ``unusual'' case is reported by solid squares in Fig.~\ref{Fig4} and corresponds to a {\em stable} square lattice obtained for potential~(\ref{U1}) with $na^2=1.44$. In this case, the imaginary mode with $\omega^2<0$ is no longer present. Additionally we verify the stability of the system in Monte Carlo simulations for temperatures below $T_c$. The second ``unusual'' case corresponds to an {\em unstable} triangular lattice and is shown by open triangles in Fig.~\ref{Fig4}. In this case there is a number of imaginary modes which make the system unstable. We have verified the stability of all lattice configurations reported in Figs.~\ref{Fig1}-\ref{Fig2}. In the case of the stable molecular lattice, the eigenmodes come in pairs and each level is twice degenerate.

The obtained results might be relevant in the context of recent observation of Rydberg atoms \cite{Saffman2009,Gatan2009,Schau2012} and dipoles \cite{PhysRevLett.97.083003,PhysRevLett.99.073002,Comparat2010} in the blockade regime. The long-range part of the interaction potential has $1/r^6$ tail while the short-range part is dominated by the Coulomb blockade. Kagomé lattice has been experimentally created by using optical trapping\cite{PhysRevX.4.021034} while here we predict spontaneous formation if the specific conditions are met. The Kagomé lattice is important for creation of topological insulators \cite{PhysRevLett.109.145301}, flat-band systems \cite{Li2018}, ferromagnetism \cite{PhysRevLett.121.096401} and Majorana models \cite{PhysRevB.100.195146}.

\section{Conclusion}

Usually, the triangular lattice is formed in the ground-state of two-dimensional classical many-body systems. Here we provide a number of specific examples of 2D systems consisting of atoms interacting via spherical repulsive monotonically decaying pair interaction potential for which the triangular lattice becomes unstable. Instead, a variety of different lattice configurations is found, including Kagomé, ``flower'', molecular and chain crystals and even square lattices. The results are obtained by using classical Monte Carlo method in combination with the annealing technique. We provide a simple interpretation of the instability in terms of the harmonic theory. The obtained results might be relevant for experiments with Rydberg atoms in the blockade regime.

\section*{Acknowlegements}

This study has been partially supported by funding from the Spanish MINECO (FIS2017-84114-C2-1-P). The Barcelona Supercomputing Center (The Spanish National Supercomputing Center - Centro Nacional de Supercomputaci\'on) is acknowledged for the provided computational facilities (RES-FI-2019-2-0033). I.L. Kurbakov was supported by RFBR 17-02-01322 grant.


\begin{thebibliography}{30}%
\makeatletter
\providecommand \@ifxundefined [1]{%
 \@ifx{#1\undefined}
}%
\providecommand \@ifnum [1]{%
 \ifnum #1\expandafter \@firstoftwo
 \else \expandafter \@secondoftwo
 \fi
}%
\providecommand \@ifx [1]{%
 \ifx #1\expandafter \@firstoftwo
 \else \expandafter \@secondoftwo
 \fi
}%
\providecommand \natexlab [1]{#1}%
\providecommand \enquote  [1]{``#1''}%
\providecommand \bibnamefont  [1]{#1}%
\providecommand \bibfnamefont [1]{#1}%
\providecommand \citenamefont [1]{#1}%
\providecommand \href@noop [0]{\@secondoftwo}%
\providecommand \href [0]{\begingroup \@sanitize@url \@href}%
\providecommand \@href[1]{\@@startlink{#1}\@@href}%
\providecommand \@@href[1]{\endgroup#1\@@endlink}%
\providecommand \@sanitize@url [0]{\catcode `\\12\catcode `\$12\catcode
  `\&12\catcode `\#12\catcode `\^12\catcode `\_12\catcode `\%12\relax}%
\providecommand \@@startlink[1]{}%
\providecommand \@@endlink[0]{}%
\providecommand \url  [0]{\begingroup\@sanitize@url \@url }%
\providecommand \@url [1]{\endgroup\@href {#1}{\urlprefix }}%
\providecommand \urlprefix  [0]{URL }%
\providecommand \Eprint [0]{\href }%
\providecommand \doibase [0]{http://dx.doi.org/}%
\providecommand \selectlanguage [0]{\@gobble}%
\providecommand \bibinfo  [0]{\@secondoftwo}%
\providecommand \bibfield  [0]{\@secondoftwo}%
\providecommand \translation [1]{[#1]}%
\providecommand \BibitemOpen [0]{}%
\providecommand \bibitemStop [0]{}%
\providecommand \bibitemNoStop [0]{.\EOS\space}%
\providecommand \EOS [0]{\spacefactor3000\relax}%
\providecommand \BibitemShut  [1]{\csname bibitem#1\endcsname}%
\let\auto@bib@innerbib\@empty
\bibitem [{\citenamefont {Kepler}(1611)}]{Kepler1611}%
  \BibitemOpen
  \bibfield  {author} {\bibinfo {author} {\bibfnamefont {J.}~\bibnamefont
  {Kepler}},\ }\href@noop {} {\emph {\bibinfo {title} {Strena, Seu, de Nive
  Sexangula}}}\ (\bibinfo  {publisher} {Godfrey Tampach},\ \bibinfo {address}
  {Frankfurt-am-Main, Germany},\ \bibinfo {year} {1611})\BibitemShut {NoStop}%
\bibitem [{\citenamefont {Kepler}(1966)}]{Kepler1966english}%
  \BibitemOpen
  \bibinfo {editor} {\bibfnamefont {J.}~\bibnamefont {Kepler}},\ ed.,\
  \href@noop {} {\emph {\bibinfo {title} {The six-cornered snowflake}}}\
  (\bibinfo  {publisher} {Clarendon Press},\ \bibinfo {address} {Oxford, UK},\
  \bibinfo {year} {1966})\ pp.\ \bibinfo {pages} {xvi + 75}\BibitemShut
  {NoStop}%
\bibitem [{\citenamefont {Hales}\ \emph {et~al.}(2015)\citenamefont {Hales},
  \citenamefont {Adams}, \citenamefont {Bauer}, \citenamefont {Dang},
  \citenamefont {Harrison}, \citenamefont {Hoang}, \citenamefont {Kaliszyk},
  \citenamefont {Magron}, \citenamefont {McLaughlin}, \citenamefont {Nguyen},
  \citenamefont {Nguyen}, \citenamefont {Nipkow}, \citenamefont {Obua},
  \citenamefont {Pleso}, \citenamefont {Rute}, \citenamefont {Solovyev},
  \citenamefont {Ta}, \citenamefont {Tran}, \citenamefont {Trieu},
  \citenamefont {Urban}, \citenamefont {Vu},\ and\ \citenamefont
  {Zumkeller}}]{arxiv1501.02155}%
  \BibitemOpen
  \bibfield  {author} {\bibinfo {author} {\bibfnamefont {T.}~\bibnamefont
  {Hales}}, \bibinfo {author} {\bibfnamefont {M.}~\bibnamefont {Adams}},
  \bibinfo {author} {\bibfnamefont {G.}~\bibnamefont {Bauer}}, \bibinfo
  {author} {\bibfnamefont {D.~T.}\ \bibnamefont {Dang}}, \bibinfo {author}
  {\bibfnamefont {J.}~\bibnamefont {Harrison}}, \bibinfo {author}
  {\bibfnamefont {T.~L.}\ \bibnamefont {Hoang}}, \bibinfo {author}
  {\bibfnamefont {C.}~\bibnamefont {Kaliszyk}}, \bibinfo {author}
  {\bibfnamefont {V.}~\bibnamefont {Magron}}, \bibinfo {author} {\bibfnamefont
  {S.}~\bibnamefont {McLaughlin}}, \bibinfo {author} {\bibfnamefont {T.~T.}\
  \bibnamefont {Nguyen}}, \bibinfo {author} {\bibfnamefont {T.~Q.}\
  \bibnamefont {Nguyen}}, \bibinfo {author} {\bibfnamefont {T.}~\bibnamefont
  {Nipkow}}, \bibinfo {author} {\bibfnamefont {S.}~\bibnamefont {Obua}},
  \bibinfo {author} {\bibfnamefont {J.}~\bibnamefont {Pleso}}, \bibinfo
  {author} {\bibfnamefont {J.}~\bibnamefont {Rute}}, \bibinfo {author}
  {\bibfnamefont {A.}~\bibnamefont {Solovyev}}, \bibinfo {author}
  {\bibfnamefont {A.~H.~T.}\ \bibnamefont {Ta}}, \bibinfo {author}
  {\bibfnamefont {T.~N.}\ \bibnamefont {Tran}}, \bibinfo {author}
  {\bibfnamefont {D.~T.}\ \bibnamefont {Trieu}}, \bibinfo {author}
  {\bibfnamefont {J.}~\bibnamefont {Urban}}, \bibinfo {author} {\bibfnamefont
  {K.~K.}\ \bibnamefont {Vu}}, \ and\ \bibinfo {author} {\bibfnamefont
  {R.}~\bibnamefont {Zumkeller}},\ }\href@noop {} {\enquote {\bibinfo {title}
  {A formal proof of the kepler conjecture},}\ } (\bibinfo {year} {2015}),\
  \Eprint {http://arxiv.org/abs/arXiv:1501.02155} {arXiv:1501.02155}
  \BibitemShut {NoStop}%
\bibitem [{\citenamefont {HALES}\ \emph {et~al.}(2017)\citenamefont {HALES},
  \citenamefont {ADAMS}, \citenamefont {BAUER}, \citenamefont {DANG},
  \citenamefont {HARRISON}, \citenamefont {HOANG}, \citenamefont {KALISZYK},
  \citenamefont {MAGRON}, \citenamefont {MCLAUGHLIN}, \citenamefont {NGUYEN},\
  and\ \citenamefont {et~al.}}]{KeplerConjecture2017}%
  \BibitemOpen
  \bibfield  {author} {\bibinfo {author} {\bibfnamefont {T.}~\bibnamefont
  {HALES}}, \bibinfo {author} {\bibfnamefont {M.}~\bibnamefont {ADAMS}},
  \bibinfo {author} {\bibfnamefont {G.}~\bibnamefont {BAUER}}, \bibinfo
  {author} {\bibfnamefont {T.~D.}\ \bibnamefont {DANG}}, \bibinfo {author}
  {\bibfnamefont {J.}~\bibnamefont {HARRISON}}, \bibinfo {author}
  {\bibfnamefont {L.~T.}\ \bibnamefont {HOANG}}, \bibinfo {author}
  {\bibfnamefont {C.}~\bibnamefont {KALISZYK}}, \bibinfo {author}
  {\bibfnamefont {V.}~\bibnamefont {MAGRON}}, \bibinfo {author} {\bibfnamefont
  {S.}~\bibnamefont {MCLAUGHLIN}}, \bibinfo {author} {\bibfnamefont {T.~T.}\
  \bibnamefont {NGUYEN}}, \ and\ \bibinfo {author} {\bibnamefont {et~al.}},\
  }\href {\doibase 10.1017/fmp.2017.1} {\bibfield  {journal} {\bibinfo
  {journal} {Forum of Mathematics, Pi}\ }\textbf {\bibinfo {volume} {5}},\
  \bibinfo {pages} {e2} (\bibinfo {year} {2017})}\BibitemShut {NoStop}%
\bibitem [{\citenamefont {Kittel}(1963)}]{KittelBook}%
  \BibitemOpen
  \bibfield  {author} {\bibinfo {author} {\bibfnamefont {C.}~\bibnamefont
  {Kittel}},\ }\href@noop {} {\emph {\bibinfo {title} {Quantum theory of
  solids}}}\ (\bibinfo  {publisher} {New York : John Wiley \& Sons, Inc},\
  \bibinfo {year} {1963})\BibitemShut {NoStop}%
\bibitem [{\citenamefont {Abrikosov}(1957)}]{Abrikosov1957}%
  \BibitemOpen
  \bibfield  {author} {\bibinfo {author} {\bibfnamefont {A.}~\bibnamefont
  {Abrikosov}},\ }\href {\doibase https://doi.org/10.1016/0022-3697(57)90083-5}
  {\bibfield  {journal} {\bibinfo  {journal} {Journal of Physics and Chemistry
  of Solids}\ }\textbf {\bibinfo {volume} {2}},\ \bibinfo {pages} {199 }
  (\bibinfo {year} {1957})}\BibitemShut {NoStop}%
\bibitem [{\citenamefont {Henkel}\ \emph {et~al.}(2010)\citenamefont {Henkel},
  \citenamefont {Nath},\ and\ \citenamefont {Pohl}}]{PhysRevLett.104.195302}%
  \BibitemOpen
  \bibfield  {author} {\bibinfo {author} {\bibfnamefont {N.}~\bibnamefont
  {Henkel}}, \bibinfo {author} {\bibfnamefont {R.}~\bibnamefont {Nath}}, \ and\
  \bibinfo {author} {\bibfnamefont {T.}~\bibnamefont {Pohl}},\ }\href {\doibase
  10.1103/PhysRevLett.104.195302} {\bibfield  {journal} {\bibinfo  {journal}
  {Phys. Rev. Lett.}\ }\textbf {\bibinfo {volume} {104}},\ \bibinfo {pages}
  {195302} (\bibinfo {year} {2010})}\BibitemShut {NoStop}%
\bibitem [{\citenamefont {Henkel}\ \emph {et~al.}(2012)\citenamefont {Henkel},
  \citenamefont {Cinti}, \citenamefont {Jain}, \citenamefont {Pupillo},\ and\
  \citenamefont {Pohl}}]{PhysRevLett.108.265301}%
  \BibitemOpen
  \bibfield  {author} {\bibinfo {author} {\bibfnamefont {N.}~\bibnamefont
  {Henkel}}, \bibinfo {author} {\bibfnamefont {F.}~\bibnamefont {Cinti}},
  \bibinfo {author} {\bibfnamefont {P.}~\bibnamefont {Jain}}, \bibinfo {author}
  {\bibfnamefont {G.}~\bibnamefont {Pupillo}}, \ and\ \bibinfo {author}
  {\bibfnamefont {T.}~\bibnamefont {Pohl}},\ }\href {\doibase
  10.1103/PhysRevLett.108.265301} {\bibfield  {journal} {\bibinfo  {journal}
  {Phys. Rev. Lett.}\ }\textbf {\bibinfo {volume} {108}},\ \bibinfo {pages}
  {265301} (\bibinfo {year} {2012})}\BibitemShut {NoStop}%
\bibitem [{\citenamefont {Angelone}\ \emph
  {et~al.}(2016{\natexlab{a}})\citenamefont {Angelone}, \citenamefont {Ying},
  \citenamefont {Mezzacapo}, \citenamefont {Masella}, \citenamefont
  {Dalmonte},\ and\ \citenamefont {Pupillo}}]{arxiv1606.04267}%
  \BibitemOpen
  \bibfield  {author} {\bibinfo {author} {\bibfnamefont {A.}~\bibnamefont
  {Angelone}}, \bibinfo {author} {\bibfnamefont {T.}~\bibnamefont {Ying}},
  \bibinfo {author} {\bibfnamefont {F.}~\bibnamefont {Mezzacapo}}, \bibinfo
  {author} {\bibfnamefont {G.}~\bibnamefont {Masella}}, \bibinfo {author}
  {\bibfnamefont {M.}~\bibnamefont {Dalmonte}}, \ and\ \bibinfo {author}
  {\bibfnamefont {G.}~\bibnamefont {Pupillo}},\ }\href@noop {} {\enquote
  {\bibinfo {title} {Non-equilibrium scenarios in cluster-forming quantum
  lattice models},}\ } (\bibinfo {year} {2016}{\natexlab{a}}),\ \Eprint
  {http://arxiv.org/abs/arXiv:1606.04267} {arXiv:1606.04267} \BibitemShut
  {NoStop}%
\bibitem [{\citenamefont {D\'{\i}az-M\'endez}\ \emph
  {et~al.}(2015)\citenamefont {D\'{\i}az-M\'endez}, \citenamefont {Mezzacapo},
  \citenamefont {Cinti}, \citenamefont {Lechner},\ and\ \citenamefont
  {Pupillo}}]{pre092052307}%
  \BibitemOpen
  \bibfield  {author} {\bibinfo {author} {\bibfnamefont {R.}~\bibnamefont
  {D\'{\i}az-M\'endez}}, \bibinfo {author} {\bibfnamefont {F.}~\bibnamefont
  {Mezzacapo}}, \bibinfo {author} {\bibfnamefont {F.}~\bibnamefont {Cinti}},
  \bibinfo {author} {\bibfnamefont {W.}~\bibnamefont {Lechner}}, \ and\
  \bibinfo {author} {\bibfnamefont {G.}~\bibnamefont {Pupillo}},\ }\href
  {\doibase 10.1103/PhysRevE.92.052307} {\bibfield  {journal} {\bibinfo
  {journal} {Phys. Rev. E}\ }\textbf {\bibinfo {volume} {92}},\ \bibinfo
  {pages} {052307} (\bibinfo {year} {2015})}\BibitemShut {NoStop}%
\bibitem [{\citenamefont {Osychenko}\ \emph {et~al.}(2011)\citenamefont
  {Osychenko}, \citenamefont {Astrakharchik}, \citenamefont {Lutsyshyn},
  \citenamefont {Lozovik},\ and\ \citenamefont {Boronat}}]{Osychenko2011}%
  \BibitemOpen
  \bibfield  {author} {\bibinfo {author} {\bibfnamefont {O.~N.}\ \bibnamefont
  {Osychenko}}, \bibinfo {author} {\bibfnamefont {G.~E.}\ \bibnamefont
  {Astrakharchik}}, \bibinfo {author} {\bibfnamefont {Y.}~\bibnamefont
  {Lutsyshyn}}, \bibinfo {author} {\bibfnamefont {Y.~E.}\ \bibnamefont
  {Lozovik}}, \ and\ \bibinfo {author} {\bibfnamefont {J.}~\bibnamefont
  {Boronat}},\ }\href {\doibase 10.1103/PhysRevA.84.063621} {\bibfield
  {journal} {\bibinfo  {journal} {Phys. Rev. A}\ }\textbf {\bibinfo {volume}
  {84}},\ \bibinfo {pages} {063621} (\bibinfo {year} {2011})}\BibitemShut
  {NoStop}%
\bibitem [{\citenamefont {Cinti}\ \emph {et~al.}(2010)\citenamefont {Cinti},
  \citenamefont {Jain}, \citenamefont {Boninsegni}, \citenamefont {Micheli},
  \citenamefont {Zoller},\ and\ \citenamefont
  {Pupillo}}]{PhysRevLett.105.135301}%
  \BibitemOpen
  \bibfield  {author} {\bibinfo {author} {\bibfnamefont {F.}~\bibnamefont
  {Cinti}}, \bibinfo {author} {\bibfnamefont {P.}~\bibnamefont {Jain}},
  \bibinfo {author} {\bibfnamefont {M.}~\bibnamefont {Boninsegni}}, \bibinfo
  {author} {\bibfnamefont {A.}~\bibnamefont {Micheli}}, \bibinfo {author}
  {\bibfnamefont {P.}~\bibnamefont {Zoller}}, \ and\ \bibinfo {author}
  {\bibfnamefont {G.}~\bibnamefont {Pupillo}},\ }\href {\doibase
  10.1103/PhysRevLett.105.135301} {\bibfield  {journal} {\bibinfo  {journal}
  {Phys. Rev. Lett.}\ }\textbf {\bibinfo {volume} {105}},\ \bibinfo {pages}
  {135301} (\bibinfo {year} {2010})}\BibitemShut {NoStop}%
\bibitem [{\citenamefont {Angelone}\ \emph
  {et~al.}(2016{\natexlab{b}})\citenamefont {Angelone}, \citenamefont
  {Mezzacapo},\ and\ \citenamefont {Pupillo}}]{PhysRevLett.116.135303}%
  \BibitemOpen
  \bibfield  {author} {\bibinfo {author} {\bibfnamefont {A.}~\bibnamefont
  {Angelone}}, \bibinfo {author} {\bibfnamefont {F.}~\bibnamefont {Mezzacapo}},
  \ and\ \bibinfo {author} {\bibfnamefont {G.}~\bibnamefont {Pupillo}},\ }\href
  {\doibase 10.1103/PhysRevLett.116.135303} {\bibfield  {journal} {\bibinfo
  {journal} {Phys. Rev. Lett.}\ }\textbf {\bibinfo {volume} {116}},\ \bibinfo
  {pages} {135303} (\bibinfo {year} {2016}{\natexlab{b}})}\BibitemShut
  {NoStop}%
\bibitem [{\citenamefont {Mekata}(2003)}]{Mekata2003}%
  \BibitemOpen
  \bibfield  {author} {\bibinfo {author} {\bibfnamefont {M.}~\bibnamefont
  {Mekata}},\ }\href {\doibase 10.1063/1.1564329} {\bibfield  {journal}
  {\bibinfo  {journal} {Physics Today}\ }\textbf {\bibinfo {volume} {56}},\
  \bibinfo {pages} {12} (\bibinfo {year} {2003})}\BibitemShut {NoStop}%
\bibitem [{\citenamefont {Cazorla}\ and\ \citenamefont
  {Boronat}(2017)}]{RevModPhys.89.035003}%
  \BibitemOpen
  \bibfield  {author} {\bibinfo {author} {\bibfnamefont {C.}~\bibnamefont
  {Cazorla}}\ and\ \bibinfo {author} {\bibfnamefont {J.}~\bibnamefont
  {Boronat}},\ }\href {\doibase 10.1103/RevModPhys.89.035003} {\bibfield
  {journal} {\bibinfo  {journal} {Rev. Mod. Phys.}\ }\textbf {\bibinfo {volume}
  {89}},\ \bibinfo {pages} {035003} (\bibinfo {year} {2017})}\BibitemShut
  {NoStop}%
\bibitem [{\citenamefont {Bedanov}\ \emph {et~al.}(1985)\citenamefont
  {Bedanov}, \citenamefont {Gadiyak},\ and\ \citenamefont
  {Lozovik}}]{pla109000289}%
  \BibitemOpen
  \bibfield  {author} {\bibinfo {author} {\bibfnamefont {V.}~\bibnamefont
  {Bedanov}}, \bibinfo {author} {\bibfnamefont {G.}~\bibnamefont {Gadiyak}}, \
  and\ \bibinfo {author} {\bibfnamefont {Y.}~\bibnamefont {Lozovik}},\ }\href
  {\doibase https://doi.org/10.1016/0375-9601(85)90617-6} {\bibfield  {journal}
  {\bibinfo  {journal} {Physics Letters A}\ }\textbf {\bibinfo {volume}
  {109}},\ \bibinfo {pages} {289 } (\bibinfo {year} {1985})}\BibitemShut
  {NoStop}%
\bibitem [{\citenamefont {Bonsall}\ and\ \citenamefont
  {Maradudin}(1977)}]{prb015001959}%
  \BibitemOpen
  \bibfield  {author} {\bibinfo {author} {\bibfnamefont {L.}~\bibnamefont
  {Bonsall}}\ and\ \bibinfo {author} {\bibfnamefont {A.~A.}\ \bibnamefont
  {Maradudin}},\ }\href {\doibase 10.1103/PhysRevB.15.1959} {\bibfield
  {journal} {\bibinfo  {journal} {Phys. Rev. B}\ }\textbf {\bibinfo {volume}
  {15}},\ \bibinfo {pages} {1959} (\bibinfo {year} {1977})}\BibitemShut
  {NoStop}%
\bibitem [{\citenamefont {Komsa}\ and\ \citenamefont
  {Krasheninnikov}(2012)}]{DFT}%
  \BibitemOpen
  \bibfield  {author} {\bibinfo {author} {\bibfnamefont {H.-P.}\ \bibnamefont
  {Komsa}}\ and\ \bibinfo {author} {\bibfnamefont {A.~V.}\ \bibnamefont
  {Krasheninnikov}},\ }\href {\doibase 10.1103/PhysRevB.86.241201} {\bibfield
  {journal} {\bibinfo  {journal} {Phys. Rev. B}\ }\textbf {\bibinfo {volume}
  {86}},\ \bibinfo {pages} {241201} (\bibinfo {year} {2012})}\BibitemShut
  {NoStop}%
\bibitem [{\citenamefont {Ashcroft}\ and\ \citenamefont
  {Mermin}(1976)}]{Ashcroft76}%
  \BibitemOpen
  \bibfield  {author} {\bibinfo {author} {\bibfnamefont {N.~W.}\ \bibnamefont
  {Ashcroft}}\ and\ \bibinfo {author} {\bibfnamefont {N.~D.}\ \bibnamefont
  {Mermin}},\ }\href@noop {} {\emph {\bibinfo {title} {Solid State Physics}}}\
  (\bibinfo  {publisher} {Holt, Rinehart, and Winston, Philadelphia},\ \bibinfo
  {year} {1976})\BibitemShut {NoStop}%
\bibitem [{\citenamefont {Urban}\ \emph {et~al.}(2009)\citenamefont {Urban},
  \citenamefont {Johnson}, \citenamefont {Henage}, \citenamefont {Isenhower},
  \citenamefont {Yavuz}, \citenamefont {Walker},\ and\ \citenamefont
  {Saffman}}]{Saffman2009}%
  \BibitemOpen
  \bibfield  {author} {\bibinfo {author} {\bibfnamefont {E.}~\bibnamefont
  {Urban}}, \bibinfo {author} {\bibfnamefont {T.~A.}\ \bibnamefont {Johnson}},
  \bibinfo {author} {\bibfnamefont {T.}~\bibnamefont {Henage}}, \bibinfo
  {author} {\bibfnamefont {L.}~\bibnamefont {Isenhower}}, \bibinfo {author}
  {\bibfnamefont {D.~D.}\ \bibnamefont {Yavuz}}, \bibinfo {author}
  {\bibfnamefont {T.~G.}\ \bibnamefont {Walker}}, \ and\ \bibinfo {author}
  {\bibfnamefont {M.}~\bibnamefont {Saffman}},\ }\href {\doibase
  10.1038/nphys1178} {\bibfield  {journal} {\bibinfo  {journal} {Nature
  Physics}\ }\textbf {\bibinfo {volume} {5}},\ \bibinfo {pages} {110} (\bibinfo
  {year} {2009})}\BibitemShut {NoStop}%
\bibitem [{\citenamefont {Gaëtan}\ \emph {et~al.}(2009)\citenamefont
  {Gaëtan}, \citenamefont {Miroshnychenko}, \citenamefont {Wilk},
  \citenamefont {Chotia}, \citenamefont {Viteau}, \citenamefont {Comparat},
  \citenamefont {Pillet}, \citenamefont {Browaeys},\ and\ \citenamefont
  {Grangier}}]{Gatan2009}%
  \BibitemOpen
  \bibfield  {author} {\bibinfo {author} {\bibfnamefont {A.}~\bibnamefont
  {Gaëtan}}, \bibinfo {author} {\bibfnamefont {Y.}~\bibnamefont
  {Miroshnychenko}}, \bibinfo {author} {\bibfnamefont {T.}~\bibnamefont
  {Wilk}}, \bibinfo {author} {\bibfnamefont {A.}~\bibnamefont {Chotia}},
  \bibinfo {author} {\bibfnamefont {M.}~\bibnamefont {Viteau}}, \bibinfo
  {author} {\bibfnamefont {D.}~\bibnamefont {Comparat}}, \bibinfo {author}
  {\bibfnamefont {P.}~\bibnamefont {Pillet}}, \bibinfo {author} {\bibfnamefont
  {A.}~\bibnamefont {Browaeys}}, \ and\ \bibinfo {author} {\bibfnamefont
  {P.}~\bibnamefont {Grangier}},\ }\href {\doibase 10.1038/nphys1183}
  {\bibfield  {journal} {\bibinfo  {journal} {Nature Physics}\ }\textbf
  {\bibinfo {volume} {5}},\ \bibinfo {pages} {115} (\bibinfo {year}
  {2009})}\BibitemShut {NoStop}%
\bibitem [{\citenamefont {Schau{\ss}}\ \emph {et~al.}(2012)\citenamefont
  {Schau{\ss}}, \citenamefont {Cheneau}, \citenamefont {Endres}, \citenamefont
  {Fukuhara}, \citenamefont {Hild}, \citenamefont {Omran}, \citenamefont
  {Pohl}, \citenamefont {Gross}, \citenamefont {Kuhr},\ and\ \citenamefont
  {Bloch}}]{Schau2012}%
  \BibitemOpen
  \bibfield  {author} {\bibinfo {author} {\bibfnamefont {P.}~\bibnamefont
  {Schau{\ss}}}, \bibinfo {author} {\bibfnamefont {M.}~\bibnamefont {Cheneau}},
  \bibinfo {author} {\bibfnamefont {M.}~\bibnamefont {Endres}}, \bibinfo
  {author} {\bibfnamefont {T.}~\bibnamefont {Fukuhara}}, \bibinfo {author}
  {\bibfnamefont {S.}~\bibnamefont {Hild}}, \bibinfo {author} {\bibfnamefont
  {A.}~\bibnamefont {Omran}}, \bibinfo {author} {\bibfnamefont
  {T.}~\bibnamefont {Pohl}}, \bibinfo {author} {\bibfnamefont {C.}~\bibnamefont
  {Gross}}, \bibinfo {author} {\bibfnamefont {S.}~\bibnamefont {Kuhr}}, \ and\
  \bibinfo {author} {\bibfnamefont {I.}~\bibnamefont {Bloch}},\ }\href
  {\doibase 10.1038/nature11596} {\bibfield  {journal} {\bibinfo  {journal}
  {Nature}\ }\textbf {\bibinfo {volume} {491}},\ \bibinfo {pages} {87}
  (\bibinfo {year} {2012})}\BibitemShut {NoStop}%
\bibitem [{\citenamefont {Vogt}\ \emph {et~al.}(2006)\citenamefont {Vogt},
  \citenamefont {Viteau}, \citenamefont {Zhao}, \citenamefont {Chotia},
  \citenamefont {Comparat},\ and\ \citenamefont
  {Pillet}}]{PhysRevLett.97.083003}%
  \BibitemOpen
  \bibfield  {author} {\bibinfo {author} {\bibfnamefont {T.}~\bibnamefont
  {Vogt}}, \bibinfo {author} {\bibfnamefont {M.}~\bibnamefont {Viteau}},
  \bibinfo {author} {\bibfnamefont {J.}~\bibnamefont {Zhao}}, \bibinfo {author}
  {\bibfnamefont {A.}~\bibnamefont {Chotia}}, \bibinfo {author} {\bibfnamefont
  {D.}~\bibnamefont {Comparat}}, \ and\ \bibinfo {author} {\bibfnamefont
  {P.}~\bibnamefont {Pillet}},\ }\href {\doibase 10.1103/PhysRevLett.97.083003}
  {\bibfield  {journal} {\bibinfo  {journal} {Phys. Rev. Lett.}\ }\textbf
  {\bibinfo {volume} {97}},\ \bibinfo {pages} {083003} (\bibinfo {year}
  {2006})}\BibitemShut {NoStop}%
\bibitem [{\citenamefont {Vogt}\ \emph {et~al.}(2007)\citenamefont {Vogt},
  \citenamefont {Viteau}, \citenamefont {Chotia}, \citenamefont {Zhao},
  \citenamefont {Comparat},\ and\ \citenamefont
  {Pillet}}]{PhysRevLett.99.073002}%
  \BibitemOpen
  \bibfield  {author} {\bibinfo {author} {\bibfnamefont {T.}~\bibnamefont
  {Vogt}}, \bibinfo {author} {\bibfnamefont {M.}~\bibnamefont {Viteau}},
  \bibinfo {author} {\bibfnamefont {A.}~\bibnamefont {Chotia}}, \bibinfo
  {author} {\bibfnamefont {J.}~\bibnamefont {Zhao}}, \bibinfo {author}
  {\bibfnamefont {D.}~\bibnamefont {Comparat}}, \ and\ \bibinfo {author}
  {\bibfnamefont {P.}~\bibnamefont {Pillet}},\ }\href {\doibase
  10.1103/PhysRevLett.99.073002} {\bibfield  {journal} {\bibinfo  {journal}
  {Phys. Rev. Lett.}\ }\textbf {\bibinfo {volume} {99}},\ \bibinfo {pages}
  {073002} (\bibinfo {year} {2007})}\BibitemShut {NoStop}%
\bibitem [{\citenamefont {Comparat}\ and\ \citenamefont
  {Pillet}(2010)}]{Comparat2010}%
  \BibitemOpen
  \bibfield  {author} {\bibinfo {author} {\bibfnamefont {D.}~\bibnamefont
  {Comparat}}\ and\ \bibinfo {author} {\bibfnamefont {P.}~\bibnamefont
  {Pillet}},\ }\href {\doibase 10.1364/JOSAB.27.00A208} {\bibfield  {journal}
  {\bibinfo  {journal} {J. Opt. Soc. Am. B}\ }\textbf {\bibinfo {volume}
  {27}},\ \bibinfo {pages} {A208} (\bibinfo {year} {2010})}\BibitemShut
  {NoStop}%
\bibitem [{\citenamefont {Nogrette}\ \emph {et~al.}(2014)\citenamefont
  {Nogrette}, \citenamefont {Labuhn}, \citenamefont {Ravets}, \citenamefont
  {Barredo}, \citenamefont {B\'eguin}, \citenamefont {Vernier}, \citenamefont
  {Lahaye},\ and\ \citenamefont {Browaeys}}]{PhysRevX.4.021034}%
  \BibitemOpen
  \bibfield  {author} {\bibinfo {author} {\bibfnamefont {F.}~\bibnamefont
  {Nogrette}}, \bibinfo {author} {\bibfnamefont {H.}~\bibnamefont {Labuhn}},
  \bibinfo {author} {\bibfnamefont {S.}~\bibnamefont {Ravets}}, \bibinfo
  {author} {\bibfnamefont {D.}~\bibnamefont {Barredo}}, \bibinfo {author}
  {\bibfnamefont {L.}~\bibnamefont {B\'eguin}}, \bibinfo {author}
  {\bibfnamefont {A.}~\bibnamefont {Vernier}}, \bibinfo {author} {\bibfnamefont
  {T.}~\bibnamefont {Lahaye}}, \ and\ \bibinfo {author} {\bibfnamefont
  {A.}~\bibnamefont {Browaeys}},\ }\href {\doibase 10.1103/PhysRevX.4.021034}
  {\bibfield  {journal} {\bibinfo  {journal} {Phys. Rev. X}\ }\textbf {\bibinfo
  {volume} {4}},\ \bibinfo {pages} {021034} (\bibinfo {year}
  {2014})}\BibitemShut {NoStop}%
\bibitem [{\citenamefont {Hauke}\ \emph {et~al.}(2012)\citenamefont {Hauke},
  \citenamefont {Tieleman}, \citenamefont {Celi}, \citenamefont
  {\"Olschl\"ager}, \citenamefont {Simonet}, \citenamefont {Struck},
  \citenamefont {Weinberg}, \citenamefont {Windpassinger}, \citenamefont
  {Sengstock}, \citenamefont {Lewenstein},\ and\ \citenamefont
  {Eckardt}}]{PhysRevLett.109.145301}%
  \BibitemOpen
  \bibfield  {author} {\bibinfo {author} {\bibfnamefont {P.}~\bibnamefont
  {Hauke}}, \bibinfo {author} {\bibfnamefont {O.}~\bibnamefont {Tieleman}},
  \bibinfo {author} {\bibfnamefont {A.}~\bibnamefont {Celi}}, \bibinfo {author}
  {\bibfnamefont {C.}~\bibnamefont {\"Olschl\"ager}}, \bibinfo {author}
  {\bibfnamefont {J.}~\bibnamefont {Simonet}}, \bibinfo {author} {\bibfnamefont
  {J.}~\bibnamefont {Struck}}, \bibinfo {author} {\bibfnamefont
  {M.}~\bibnamefont {Weinberg}}, \bibinfo {author} {\bibfnamefont
  {P.}~\bibnamefont {Windpassinger}}, \bibinfo {author} {\bibfnamefont
  {K.}~\bibnamefont {Sengstock}}, \bibinfo {author} {\bibfnamefont
  {M.}~\bibnamefont {Lewenstein}}, \ and\ \bibinfo {author} {\bibfnamefont
  {A.}~\bibnamefont {Eckardt}},\ }\href {\doibase
  10.1103/PhysRevLett.109.145301} {\bibfield  {journal} {\bibinfo  {journal}
  {Phys. Rev. Lett.}\ }\textbf {\bibinfo {volume} {109}},\ \bibinfo {pages}
  {145301} (\bibinfo {year} {2012})}\BibitemShut {NoStop}%
\bibitem [{\citenamefont {Li}\ \emph {et~al.}(2018)\citenamefont {Li},
  \citenamefont {Zhuang}, \citenamefont {Wang}, \citenamefont {Feng},
  \citenamefont {Gao}, \citenamefont {Xu}, \citenamefont {Hao}, \citenamefont
  {Wang}, \citenamefont {Zhang}, \citenamefont {Wu}, \citenamefont {Dou},
  \citenamefont {Chen}, \citenamefont {Hu},\ and\ \citenamefont {Du}}]{Li2018}%
  \BibitemOpen
  \bibfield  {author} {\bibinfo {author} {\bibfnamefont {Z.}~\bibnamefont
  {Li}}, \bibinfo {author} {\bibfnamefont {J.}~\bibnamefont {Zhuang}}, \bibinfo
  {author} {\bibfnamefont {L.}~\bibnamefont {Wang}}, \bibinfo {author}
  {\bibfnamefont {H.}~\bibnamefont {Feng}}, \bibinfo {author} {\bibfnamefont
  {Q.}~\bibnamefont {Gao}}, \bibinfo {author} {\bibfnamefont {X.}~\bibnamefont
  {Xu}}, \bibinfo {author} {\bibfnamefont {W.}~\bibnamefont {Hao}}, \bibinfo
  {author} {\bibfnamefont {X.}~\bibnamefont {Wang}}, \bibinfo {author}
  {\bibfnamefont {C.}~\bibnamefont {Zhang}}, \bibinfo {author} {\bibfnamefont
  {K.}~\bibnamefont {Wu}}, \bibinfo {author} {\bibfnamefont {S.~X.}\
  \bibnamefont {Dou}}, \bibinfo {author} {\bibfnamefont {L.}~\bibnamefont
  {Chen}}, \bibinfo {author} {\bibfnamefont {Z.}~\bibnamefont {Hu}}, \ and\
  \bibinfo {author} {\bibfnamefont {Y.}~\bibnamefont {Du}},\ }\href {\doibase
  10.1126/sciadv.aau4511} {\bibfield  {journal} {\bibinfo  {journal} {Science
  Advances}\ }\textbf {\bibinfo {volume} {4}},\ \bibinfo {pages} {eaau4511}
  (\bibinfo {year} {2018})}\BibitemShut {NoStop}%
\bibitem [{\citenamefont {Lin}\ \emph {et~al.}(2018)\citenamefont {Lin},
  \citenamefont {Choi}, \citenamefont {Zhang}, \citenamefont {Qin},
  \citenamefont {Yi}, \citenamefont {Wang}, \citenamefont {Li}, \citenamefont
  {Wang}, \citenamefont {Zhang}, \citenamefont {Sun}, \citenamefont {Wei},
  \citenamefont {Zhang}, \citenamefont {Guo}, \citenamefont {Lu}, \citenamefont
  {Cho}, \citenamefont {Zeng},\ and\ \citenamefont
  {Zhang}}]{PhysRevLett.121.096401}%
  \BibitemOpen
  \bibfield  {author} {\bibinfo {author} {\bibfnamefont {Z.}~\bibnamefont
  {Lin}}, \bibinfo {author} {\bibfnamefont {J.-H.}\ \bibnamefont {Choi}},
  \bibinfo {author} {\bibfnamefont {Q.}~\bibnamefont {Zhang}}, \bibinfo
  {author} {\bibfnamefont {W.}~\bibnamefont {Qin}}, \bibinfo {author}
  {\bibfnamefont {S.}~\bibnamefont {Yi}}, \bibinfo {author} {\bibfnamefont
  {P.}~\bibnamefont {Wang}}, \bibinfo {author} {\bibfnamefont {L.}~\bibnamefont
  {Li}}, \bibinfo {author} {\bibfnamefont {Y.}~\bibnamefont {Wang}}, \bibinfo
  {author} {\bibfnamefont {H.}~\bibnamefont {Zhang}}, \bibinfo {author}
  {\bibfnamefont {Z.}~\bibnamefont {Sun}}, \bibinfo {author} {\bibfnamefont
  {L.}~\bibnamefont {Wei}}, \bibinfo {author} {\bibfnamefont {S.}~\bibnamefont
  {Zhang}}, \bibinfo {author} {\bibfnamefont {T.}~\bibnamefont {Guo}}, \bibinfo
  {author} {\bibfnamefont {Q.}~\bibnamefont {Lu}}, \bibinfo {author}
  {\bibfnamefont {J.-H.}\ \bibnamefont {Cho}}, \bibinfo {author} {\bibfnamefont
  {C.}~\bibnamefont {Zeng}}, \ and\ \bibinfo {author} {\bibfnamefont
  {Z.}~\bibnamefont {Zhang}},\ }\href {\doibase 10.1103/PhysRevLett.121.096401}
  {\bibfield  {journal} {\bibinfo  {journal} {Phys. Rev. Lett.}\ }\textbf
  {\bibinfo {volume} {121}},\ \bibinfo {pages} {096401} (\bibinfo {year}
  {2018})}\BibitemShut {NoStop}%
\bibitem [{\citenamefont {Li}\ \emph {et~al.}(2019)\citenamefont {Li},
  \citenamefont {Lantagne-Hurtubise},\ and\ \citenamefont
  {Franz}}]{PhysRevB.100.195146}%
  \BibitemOpen
  \bibfield  {author} {\bibinfo {author} {\bibfnamefont {C.}~\bibnamefont
  {Li}}, \bibinfo {author} {\bibfnamefont {E.}~\bibnamefont
  {Lantagne-Hurtubise}}, \ and\ \bibinfo {author} {\bibfnamefont
  {M.}~\bibnamefont {Franz}},\ }\href {\doibase 10.1103/PhysRevB.100.195146}
  {\bibfield  {journal} {\bibinfo  {journal} {Phys. Rev. B}\ }\textbf {\bibinfo
  {volume} {100}},\ \bibinfo {pages} {195146} (\bibinfo {year}
  {2019})}\BibitemShut {NoStop}%
\end{thebibliography}
\ifx \undefined \booktitle \def \booktitle#1{{{\em #1}}} \fi\ifx \undefined
  \mathbb \def \mathbb #1{{\bf #1}}\fi

\end{document}